\documentclass[aps,prl,preprint,groupedaddress,floatfix]{revtex4}
\usepackage{graphicx}% Include figure files
\usepackage{dcolumn}% Align table columns on decimal point
\usepackage{bm}% bold math

\newcommand{\Rey}{\mbox{\it Re}}           % Reynolds number
\newcommand{\fmode}{\tilde{\bf u}}         % Floquet mode
\newcommand{\base}{{\bf U}}                % Base flow
\newcommand{\perb}{{\bf u}'}               % Perturbation
\newcommand{\Karman}{von~K\'{a}rm\'{a}n }

\begin{document}

\bibliographystyle{prsty}

%
%==================== TITLE ====================
%

\title{Confined three-dimensional stability analysis of the cylinder wake}

\author{Dwight Barkley}
\email[]{barkley@maths.warwick.ac.uk}
\homepage[]{www.maths.warwick.ac.uk/~barkley}
%\thanks{}
%\altaffiliation{}
\affiliation{Mathematics Institute, University of Warwick, 
Coventry CV4 7AL, United Kingdom}

\date{\today}

%
%==================== ABSTRACT ====================
%
\begin{abstract}
A spatially confined stability analysis is reported for the cylinder wake at
Reynolds numbers 190 and 260.  The two three-dimensional instabilities at
these Reynolds numbers are shown to be driven by the flow just behind the
cylinder.
\end{abstract}


\maketitle

%==================== INTRODUCTION ====================

Consider the \Karman vortex street generated by flow past an infinitely long
circular cylinder. This flow is time periodic and two dimensional for Reynolds
numbers between approximately 47 and
189~\cite{Jackson87,Provansal87,Williamson88,Miller94}.  (The Reynolds number
is $\Rey \equiv U_\infty d /\nu$, where $U_\infty$ is the fluid velocity far
from the cylinder, $d$ is the cylinder diameter, and $\nu$ is the kinematic
viscosity.)  At Reynolds number $189$ the 2D vortex street becomes three
dimensionally unstable
\cite{Williamson88,Miller94,Barkley96,Henderson96,Williamson96b}.  A numerical
stability analysis of the flow up to Reynolds number 300 has determined two
separate bands of linearly unstable modes \cite{Barkley96}. The first appears
at $Re=189$ with a spanwise wavelength of $4$ cylinder diameters and the
second appears at $Re=260$ with a spanwise wavelength of $0.8$ diameters.
These linear instabilities are related to 3D shedding modes known as mode A
and mode B, first observed experimentally by Williamson \cite{Williamson88}.

In this paper I revisit the stability analysis of the vortex street. I show
that small regions of the full flow just behind the cylinder are responsible
for the 3D linear instabilities despite the fact that the actual linear modes
extend many cylinder diameters downstream of the cylinder.  This is important
for two separate reasons. The first is that this limits the regions which
should be analyzed either to clarify instability mechanisms or to suppress the
instabilities if so desired.  The second reason is that the stability analysis
of small flow regions is computationally very efficient compared with a
stability analysis of the full flow field. Thus this approach provides a
method for quickly obtaining approximate stability information about a complex
flow.

\begin{figure}
\begin{center}
\includegraphics[width=3.2in]{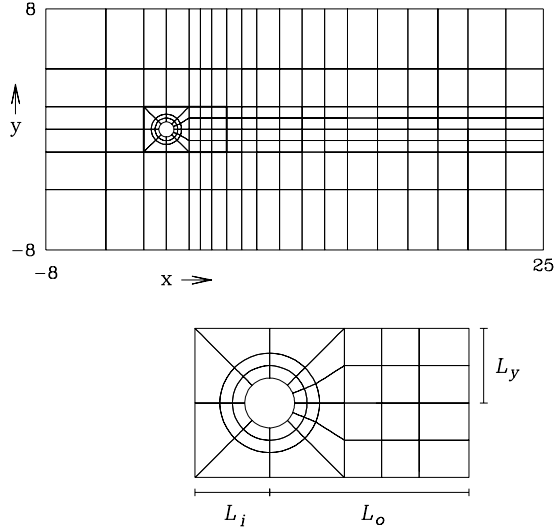}
\end{center}
\caption{{} Spectral-element computational mesh.  The base flow is computed on
the full domain and stability computations are performed on the full domain and
on a variety of subdomains. The subdomain outlined in bold and shown as an
enlargement has inflow length $L_i=1.5$, outflow length $L_o=4$, and
cross-stream height $L_y=1.5$.  }
\label{fig:mesh} 
\end{figure}

The computational approach is described fully elsewhere \cite{Barkley96}.  I
summarize briefly the main points with focus on the important aspects for the
current work.  First direct simulations of the incompressible Navier--Stokes
equations are used to obtain 2D, time-periodic wake flows: $\base(x,y,t+T) =
\base(x,y,t)$ where $T$ is the wake period. A spectral-element method is
employed on a computational domain shown in Fig.~\ref{fig:mesh}.  The length
scale is the cylinder diameter.  The size of the computational domain depends
on the Reynolds number and the accuracy required \cite{Barkley96}. The domain
shown gives sufficiently accurate results for the purposes of this work.

The boundary conditions imposed for the computation of $\base$ are as follows.
On the top, bottom, and left boundaries uniform flow is imposed $\base = (U,V)
= (1,0)$.  At the right edge a (Neumann) outflow boundary condition for the
velocity and pressure $P$ is used:
\begin{equation}
\label{bcs}
	\partial U/\partial x = 0, \quad
	\partial V/\partial x = 0, \quad 
	                    P = 0. 
\end{equation}
No-slip conditions are imposed at the cylinder surface.

The next step is a Floquet stability analysis of 3D disturbances to the 2D
wake. This analysis is based on the evolution of infinitesimal 3D
perturbations $\perb(x,y,z,t)$ of the 2D flow $\base(x,y,t)$. The equation for
such a perturbation is obtained by linearizing the Navier--Stokes equations
about $\base$:
%
% equation
%
\begin{equation}
\label{eqn:linear}
\frac{\partial{\bf u}'}{\partial t} =
     -(\base\cdot\nabla)\perb - (\perb\cdot\nabla)\base
     -\frac{1}{\rho}\nabla p' + \frac{1}{\Rey}\nabla^2\perb,
\end{equation}
where $p'$ is the perturbation to the pressure that enforces $\nabla\cdot\perb
= 0$.  

In this work I shall primarily consider Eq.~(\ref{eqn:linear}) posed on
subdomains such as that illustrated in Fig.~\ref{fig:mesh}. These subdomains
are characterized by their inflow and outflow lengths $L_i$ and $L_o$,
measured with respect to the cylinder center, and their cross-flow range $-L_y
\le y \le L_y$.  The base flow $\base$ appearing in Eq.~(\ref{eqn:linear}) is
obtained by restricting the base flow from the full domain to the appropriate
subdomain.

Boundary conditions on ${\bf u}'$ are required by Eq.~(\ref{eqn:linear}).
Consider the standard case in which ${\bf u}'$ is computed on the same domain
as $\base$. Then the requirement that $\base + \perb$ satisfy the same
boundary conditions as $\base$ gives homogeneous boundary conditions on ${\bf
u}'$: ${\bf u}'= (0,0,0)$ on all domain boundaries except the outflow boundary
where ${\bf u}'$ satisfies the analogy of Eq.~(\ref{bcs}).  One can view these
boundary conditions as the requirement that the perturbation be zero at the
inflow and lateral boundaries of the domain but that perturbations can advect
out of the domain.  Based on this reasoning we impose these homogeneous
boundary conditions when computing ${\bf u}'$ on subdomains. We are
interested in examining perturbations ${\bf u}'$ which are zero on all
boundaries of the subdomain except the outflow boundary.  Perturbations are
allowed to advect out of the subdomain.

From the linearized equations the spectrum of Floquet multipliers $\mu$ and
corresponding Floquet modes $\fmode$ can be found \cite{Barkley96}.
Exponentially growing perturbations correspond to multipliers outside the unit
circle in the complex plane ($|\mu|>1$).  Because the geometry is homogeneous
in the spanwise direction, Floquet modes decouple in spanwise wavenumber
$\beta$.  The Floquet multipliers are thus computed as a function of $\beta$
on a number of subdomains for $\Rey =190$ and $\Rey = 260$, just above the
critical Reynolds numbers for the 3D instabilities.

%\section{Results}

\begin{figure}
\begin{center}
\includegraphics[width=3.0in]{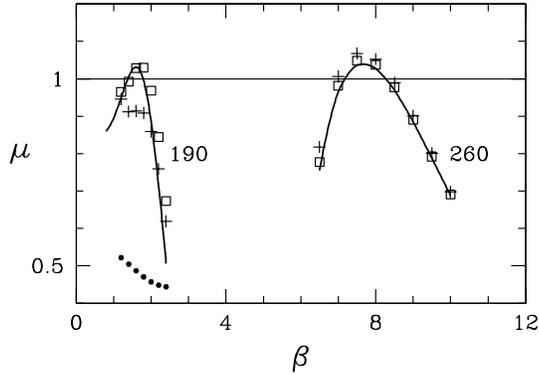}
\end{center}
\caption{{}Summary of Floquet multipliers at $\Rey =190$ and $\Rey = 260$ (as
labeled) for $\beta$ near the critical value in each case.  Solid curves are
from computations on the full domain. Points are from subdomains with
$L_i=0$, $L_y=1.5$, and $L_o=3$ (squares), $L_o=2.25$ (crosses), and $L_o=1.5$
(dots).  The multipliers for $\Rey=260$, $L_o=1.5$ are off the scale of the
figure.  }
\label{fig:mults} 
\end{figure}

\begin{figure}
\begin{center}
\includegraphics[width=3.2in]{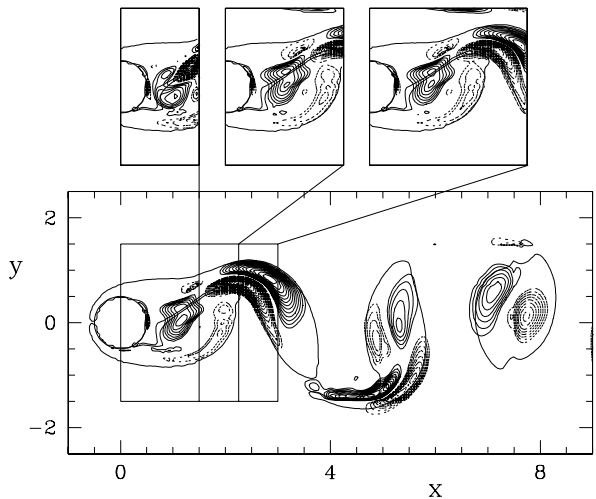}
\end{center}
\caption{{}Floquet modes at $\Rey=190$ with $\beta = 1.6$. The main figure
shows (a portion of) the mode computed on the full domain. Insets show modes
computed on the subdomains indicated. (The inflow boundary $L_i=0$ and the
lateral boundaries $L_y=1.5$ are the same for all subdomains.)  All modes are
plotted as spanwise vorticity contours at the same fixed time. The spanwise
vorticity contours $\pm 0.4$ of the base flow are also plotted.  }
\label{fig:R190} 
\end{figure}

\begin{figure}
\begin{center}
\includegraphics[width=3.2in]{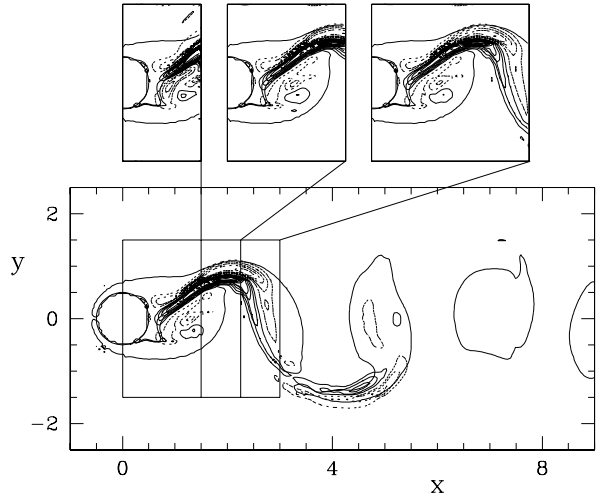}
\end{center}
\caption{{}Floquet modes at $\Rey=260$ with $\beta = 7.7$ using the same
conventions as Fig.~\ref{fig:R190}}
\label{fig:R260} 
\end{figure}

Figure~\ref{fig:mults} summarizes the study in terms of multiplier spectra
while Figs.~\ref{fig:R190}-\ref{fig:R260} shows representative Floquet modes.
A large number of subdomains have been studied but only representative cases
are shown near the minimum dimensions necessary to capture the 3D
instabilities.

Figure~\ref{fig:mults} primarily demonstrates the effect of outflow length
$L_o$ on the Floquet multiplier spectrum.  The inflow length and cross-stream
height are fixed at minimal values of $L_i=0$ and $L_y=1.5$.  One sees that
$L_o=3$ is sufficient to capture the destabilizing Floquet mode. Increasing
$L_o$ further has minimal effect.  For $\Rey=190$, decreasing $L_o$ to $2.25$
causes a substantial change to the spectrum -- the subdomain is evidently too
small to capture all the relevant flow features which drive the instability.
In the case of $\Rey=260$, decreasing $L_o$ to $2.25$ causes only a small
change to the spectrum and the subdomain is essentially still large enough to
capture relevant flow features.  Decreasing the outflow length to $L_o = 1.5$
causes the spectrum in both cases to deviate wildly from the correct behavior
(at $\Rey=260$ the multipliers become complex with large magnitude and cannot
be plotted on the scale of Fig.~\ref{fig:mults}).
Figs.~\ref{fig:R190}-\ref{fig:R260} shows just how well the Floquet modes are
captured by the subdomain computations even though the Floquet mode on the
full domain extends far downstream of the cylinder \cite{Barkley96}.

Increasing the inflow length $L_i$ has some effect on the spectrum of
multipliers in the case of $\Rey=190$. However, the effect is very small at
the critical wavenumber where the multiplier branch has largest
magnitude. Increasing the inflow length has almost no effect in the case
$\Rey=260$.  Increasing the cross-stream height has very little effect in
either case.  It is possible, in fact, that the cross-stream height could be
reduced slightly from that used here without compromising the accuracy of the
multipliers. 

It is reasonable to conclude that the physical region responsible for the
instability is localized within the subdomain in those cases where the Floquet
multipliers and Floquet modes from subdomains are close to those from the full
domain.  This localization of the instability is similar to the transition
from steady flow to vortex shedding which takes place at the primary wake
instability. In that case there is a local, absolutely unstable region
extending approximately three to four diameters downstream
\cite{Hannemann89,Oertel90}, very similar to the scales observed here.  In
fact the current computational analysis is similar in spirit to that of
Hannemann and Oertel \cite{Hannemann89} and can be thought of as situated
between the local analysis of a parallel wake profile and the global analysis
previously reported for the secondary instability to three dimensionality
\cite{Barkley96}.

Finally, it is worth considering the stability of the downstream portion of
the flow.  In this case the base flow is projected onto the subdomain with
$2.25 \le x \le 25$ and $-4 \le y \le 4$.  Both at $\Rey=190$ and $\Rey=260$
the flow in this downstream region is found to be very stable. The maximum
modulus of the Floquet multipliers are $|\mu| \simeq 0.28$ at $\Rey=190$,
$\beta=1.6$ and $|\mu| \simeq 0.19$ at $\Rey=260$, $\beta=7.5$.

In this paper I have shown how a local Floquet stability analysis of small
regions of the wake just behind a cylinder are sufficient to capture both 3D
instabilities observed in wake transition.  The specific results are of
interest for future attempts to understand better the physical mechanisms of
3D transition \cite{Williamson96b,Brede96,Persillon98,Leweke98,Thompson01}
because the results set limits to the regions which could drive the
instabilities.  The general method is of interest for future computational
studies of flow instabilities.

Much of this work was performed while visiting the Courant Institute of
Mathematical Sciences, NYU.  I thank members of the Institute for their
hospitality.

\bibliography{pre4}
% Included from pre4.bbl

%\begin{thebibliography}{10}

%\bibitem[\dagger]{byline}  email:barkley@maths.warwick.ac.uk.

%\end{thebibliography}

\bigskip
\bigskip

\end{document}